\documentclass{aa}
\usepackage{xymtex}
\usepackage{newtxtext,newtxmath}
\usepackage{graphicx} 
\usepackage{subcaption}
\usepackage{amsmath} 
\usepackage{xcolor}
\usepackage{float}
\usepackage{caption}
\usepackage{placeins}
\usepackage{amsmath}
\usepackage{lipsum}
\usepackage{multicol}
\usepackage{soul}
\usepackage[T1]{fontenc}
\usepackage{comment}

\def\refjnl#1{{\rm#1}}
\def\apjl{\refjnl{ApJ}}                % Astrophysical Journal, Letters
\def\apjs{\refjnl{ApJS}}               % Astrophysical Journal, Supplement
\def\ao{\refjnl{Appl.~Opt.}}           % Applied Optics
\def\aap{\refjnl{A\&A}}                % Astronomy and Astrophysics
\def\be{\begin{equation}} 
\def\ee{\end{equation}}

\def\msun{{\msun}}

\def\gsim{\lower.5ex\hbox{\gtsima}} 
\def\lsim{\lower.5ex\hbox{\ltsima}} \def\gtsima{$\; \buildrel > \over 
\sim \;$} \def\ltsima{$\; \buildrel < \over \sim \;$} \def\prosima{$\; 
\buildrel \propto \over \sim \;$} \def\gsim{\lower.5ex\hbox{\gtsima}} 
\def\lsim{\lower.5ex\hbox{\ltsima}} 
\def\simgt{\lower.5ex\hbox{\gtsima}} 
\def\simlt{\lower.5ex\hbox{\ltsima}} 
\def\simpr{\lower.5ex\hbox{\prosima}}   
  
 \def\gtsima{$\; \buildrel > \over \sim \;$} 
\def\ltsima{$\; \buildrel < \over \sim \;$} 
\def\gsim{\lower.5ex\hbox{\gtsima}} 
\def\lsim{\lower.5ex\hbox{\ltsima}} 
\def\simgt{\lower.5ex\hbox{\gtsima}} 
\def\simlt{\lower.5ex\hbox{\ltsima}} 
\def\simpr{\lower.5ex\hbox{\prosima}}

\def\E3{{\cal E}_{\rm g}^{III}} 
\def\msun{\rm M_\odot}

\def\mbh{\rm M_{bh}}

\def\ms{M_*}
\def\mbh{M_{\rm BH}}
\def\Z*{Z_*}
\def\L*{L_*}

\def\fej{f_*^{\rm ej}}
\def\feff{f_*^{\rm eff}}

\makeatletter
\renewcommand*\aa@pageof{, page \thepage{} of \pageref*{LastPage}}
\makeatother

\begin{document} 

   \title{Exploring a primordial solution for early black holes detected with JWST}

   \author{Pratika Dayal
          \inst{1}
                  }

   \institute{Kapteyn Astronomical Institute, University of Groningen, PO Box 800, 9700 AV Groningen, The Netherlands\\
              \email{p.dayal@rug.nl}
                            }

% **********************

  \abstract
   {  }
   {With its rest-frame optical sensitivity, the {\it James Webb} Space Telescope (JWST) has unearthed black holes as massive as $10^{6.2-8.1}\msun$ at redshifts of $z \sim 8.5-10.6$. In addition to these unexpectedly high masses, many systems show unexpectedly high ratios of the black hole to stellar mass of $\mbh/\ms \gsim 30\%$ at these early epochs. This poses a crucial challenge for theoretical models.
 }
    {We collated data for all of the black holes that were confirmed with the JWST (through spectroscopy, X-rays, or high-ionization emission lines). Using analytic calculations, we explored the combination of astrophysical seeding mechanisms and Eddington accretion rates that can explain the observed objects. We then appeal to cosmological primordial black hole (PBH) seeds and show that these present an alternative path for the seeding of early structures and their baryonic contents.  }
   {When we assume seeding (via astrophysical means) at a redshift of $z_{\rm seed}=25$ and continuous accretion, all of the black holes we studied can analytically either be explained through super-Eddington accretion (at an Eddington fraction of $f_{\rm Edd}\lsim 2.1 $) onto low-mass seeds ($100\msun$) or through Eddington-limited accretion onto high-mass seeds ($10^5\msun$). The upper limit at which we assume a primordial origin for all of these black holes yields a continuous primordial black hole mass function (between $10^{-5.25}$ and $10^{3.75}\msun$) and a fractional PBH value $\lsim 10^{-12}$. This agrees well with observational constraints. We then show that PBHs can seed a halo around themselves and assemble their baryonic (gas and stellar contents) starting at the redshift of matter-radiation equality ($z \sim 3400$). We were able to reproduce the observed stellar and black hole masses for two of the highest-redshift black holes (UHZ1 and GHZ9 at $z \sim 10.3$) with the same parameters as those that govern star formation, black hole accretion, and their feedbacks.  Exploring a wide swathe of model parameter space for GHZ9, we find ratios of black hole to stellar mass between $0.1-1.86$. This means that in some cases (of high supernova feedback), the black hole grows to be more massive than the stellar mass of its host halo. This is an attractive alternative to seeding these puzzling early systems.}
   {}

   \maketitle

% *************************************************************
\section{Introduction}
\label{sec_intro}
% *************************************************************
With its unparalleled sensitivity, the {\it James Webb} Space Telescope (JWST) has been crucial in shedding light on the black hole population in the first billion years of the Universe. These black holes comprise both intrinsically faint \citep{harikane2023bh, maiolino2023b, maiolino2024, juodzbalis2024} and heavily reddened compact sources, the so-called little red dots \citep[LRDs; e.g.][]{labbe2023, labbe2023_nat, furtak2023_phot, furtak2024, kokorev2023, matthee2023, greene2024, kokorev2024}. These early observations have given rise to a number of tantalising issues, for instance {\it (i)} clearly broadened Balmer lines have been used to infer black hole masses ranging between $10^{7-8.6}\msun$ at redshifts $z \sim 6-8.5$ \citep{kokorev2023, harikane2023bh, maiolino2023b, furtak2024,juodzbalis2024,greene2024}. Additionally, high-ionization lines \citep{maiolino2024} and X-ray detections \citep{bogdan2024, kovacs2024} have been used to infer black hole masses as high as $10^{6.2-7.9}\msun$ as early as $z \sim 10.3-10.6$. These high masses pose a crucial challenge for theoretical models of black hole seeding and growth due to the lack of cosmic time available for their assembly \citep[see reviews by e.g.][]{inayoshi2020, fan2023}. {\it (ii)} Another issue is an over-abundance of black holes in the first billion years \citep{harikane2023bh, maiolino2023b, greene2024}; and {\it (iii)} in contrast to local relations that indicate ratios of black hole ($\mbh$) to stellar mass ($\ms$) of $\lsim 0.3\%$ for high-mass ellipticals \citep[e.g.][]{volonteri2016} or even lower than $0.01\%$ \citep[e.g.][]{suh2020}, many of these early systems show unprecedentedly high ratios of $\mbh/\ms\gsim 30\%$ \citep{kokorev2023,juodzbalis2024,bogdan2024, kovacs2024} that deviate from local relations at the $>3-\sigma$ level \citep{pacucci2023}. Interestingly, however, it has been shown that many of these black holes are compatible with local $\mbh$-velocity dispersion or $\mbh$-dynamical mass relations, suggesting that while baryons might exist in the correct number in the host haloes, they are inefficient at forming stars \citep[e.g.][]{maiolino2023b}. 

A number of works, however, have urged caution in taking black hole and stellar masses at face value. Drawing a parallel with ultraluminous X-ray sources (ULXs), \citet{king2024} warned that the emission from these sources could be beamed, which would severely over-estimate the inferred luminosity. They also cautioned that standard virial indicators cannot be used because the line velocity width broadening could be dominated by outflows in un-virialised broad-line regions (BLRs). Using BLR models in different accretion regimes, \citet{lupi2024} showed that in many cases, super-Eddington accretion is preferred over the standard Shakura \& Sunyaev \citep{shakura1973} accretion disk, yielding black hole masses that are lower by an order of magnitude than those inferred by applying standard local relations to single-epoch broad lines \citep[from e.g.][] {greene2007,vestergaard2009,reines2015}. The offset of the inferred high-redshift $\mbh-\ms$ relation from local relations might be explained by a combination of selection biases (e.g. finite detection limits and broad-line detection requirements) and measurement uncertainties \citep{li2024}, and only over-massive black holes may be able to outshine their hosts and make them detectable by the JWST \citep{volonteri2023}. 

Observational issues notwithstanding, these puzzles have naturally prompted a flurry of research exploring the variety of black hole seeding and growth mechanisms allowed. In terms of {astrophysical} black hole seeds, low-mass seeds with a mass of $\sim 10^2\msun$ can be created by the collapse of metal-free (Population III) stars in mini-haloes at high redshifts \citep[e.g.][]{carr1984, abel2002, bromm2002}. Intermediate-mass seeds ($\sim 10^{3-4} \msun$) can form in dense, massive stellar clusters through a number of pathways \citep[see e.g. Sect. 2.3.1][]{amaro-seoane2023}, including dynamical interactions \citep[e.g.][]{devecchi2009}, the runaway merger of stellar mass black holes \citep[e.g.][]{belczynski2002}, or the growth of stellar mass black holes in conjunction with mergers \citep[e.g.][]{leigh2013,Alexander2014}. Finally, high-mass seeds, the so-called direct-collapse black holes (DCBHs; $\sim 10^{5} \msun$), can form via supermassive star formation \citep[e.g.][]{loeb-rasio1994,begelman2006,habouzit2016}, although this mechanism produces fewer viable seeds \citep{dayal2019}. Theoretical works allow a range of solutions for the existence of such massive black holes at early epochs so far, including requiring high-mass seeding mechanisms \citep{bogdan2024, kovacs2024, maiolino2023b, natarajan2024}, extremely efficient accretion, and merger-driven growth of either low-mass \citep{furtak2024} or intermediate-mass seeds \citep{dayal2024}, or high-mass seeds that grow through mergers with low-mass seeds and/or efficient accretion \citep{bhowmick2024}. High- or intermediate- to low-mass seeds subjected to episodic super-Eddington accretion phases are another option \citep{schneider2023, maiolino2024}, although the sustainability and effectiveness of super-Eddington accretion remains unclear \citep[e.g.][]{regan2019,massonneau2023}. It has also been postulated that the dense and dust-rich environments of LRDs, in particular, could enable rapidly spinning black holes that are maintained by prolonged mass accretion, which might result in over-massive black holes at these early epochs \citep{inayoshi2024}. Theoretical analyses have also been used to demonstrate the need for both low- and high-mass seeding mechanisms \citep{fragione2023}, suggesting a continuum of seed masses rather than a bimodal distribution \citep{regan2024}.

However, {cosmology} provides an additional tantalising seeding mechanism in terms of primordial black holes \citep[PBHs;][]{hawking1971, carr1974}. We refer to excellent reviews \citep[e.g.][]{carr2005, carr-green2024} that describe in detail how PBHs could be generated by a number of mechanisms, including collapse from scale-invariant fluctuations (or cosmic strings), by cosmological phase transitions, or by a temporary softening of the equation of state in the early Universe. We draw inspiration from a number of previous works that described that the Coulomb effect of a single black hole can generate an initial density fluctuation through the seed effect, which can grow through gravitational instability to bind (dark matter) mass around itself. This means that individual PBHs can essentially act as `seeds of structure formation \citep{hoyle1966, ryan1972, carr-rees1984, mack2007, carr-silk2018, carr2020}. These PBHs have already been used to provide solutions for dwarf galaxy anomalies \citep{silk2017}, act as the seeds of high-redshift massive structures and massive black holes \citep{mack2007, carr-silk2018, cappelluti2022, liu2023}, and to explain the excesses seen in the cosmic X-ray background \citep{ziparo2022} and radio wave background \citep{mittal2022}. 

In this work, our aim is to determine whether PBHs that assemble (isolated) haloes around themselves offer a viable pathway for generating the massive black holes detected by the JWST. We start by collating data on all of the spectroscopically confirmed black holes detected by the JWST and their physical properties in Sect. \ref{sec_model}. Assuming a seed redshift of $z_{\rm seed}=25$ appropriate for astrophysical origins, we explore the seed masses (low to high) and growth (Eddington limited to super-Eddington accretion) mechanisms required to explain this data. For black holes that cannot be explained by different seed masses and Eddington-limited growth mechanisms, we apply cosmology and infer the PBH masses at the epoch of matter-radiation equality ($z=3400$), which are used to construct the PBH mass function. We ignore any growth of PBHs before this era because this is beyond the scope of this work. Finally, starting at $z=3400$, we present an illustrative formalism of how these seeds might grow a halo around themselves, in addition to building their gas content and stellar components yielding values of $\mbh/\ms \sim 0.1-1.86$, in Sect. \ref{sec_assembly}. We discuss the role of the model free parameters in Sect. \ref{sec_parex} before we conclude in Sect. \ref{sec_conc}. 

We adopt a $\Lambda$ cold dark matter model with dark energy, dark matter, and baryonic densities in units of the critical density as $\Omega_{\Lambda}= 0.673$, $\Omega_{m}= 0.315$, and $\Omega_{b}= 0.049$, respectively, a Hubble constant $H_0=100\, h\,{\rm km}\,{\rm s}^{-1}\,{\rm Mpc}^{-1}$ with $h=0.673$, a spectral index $n=0.96$, and a normalisation $\sigma_{8}=0.81$ \citep[][]{planck2020}. Throughout this work, we use a Salpeter initial mass function \citep[IMF;][]{salpeter1955} between $0.1-100 \msun$ for the mass distribution of stars in a newly formed stellar population. 

\begin{table*}
  \caption{Black holes confirmed by the JWST at $z \gsim 4$.}
\centering
    \begin{tabular}{|c|c|c|c|c|c|c|c|c|}
    \hline
   No. & ID & Redshift & Log $\frac{M_{BH}}{\msun}$ & Signature & Log$\frac{{\rm N_{LF}}}{{\rm cMpc^3}}$ & Log$\frac{{\rm N_i}}{{\rm cMpc^3}}$  & Reference & Log $\frac{M_{\rm PBH}}{\msun}$ \\
    \hline
    1 & GHZ9 & $10.4$ & $7.9^{+0.16}_{-0.22}$ & X-rays & -5.27 & -5.57 & \citet{kovacs2024} & 3.9 \\
    2 &UHZ1 & $10.3$ & $7.5^{+0.5}_{-0.5}$ & X-rays & -5.27 & -5.57 & \citet{bogdan2024} & 3.4 \\
    3 &Abell2744-QSO1 & $7.04$ & $7.47^{+0.22}_{-0.17}$ & H$\beta$ & -4.13 & -4.44 & \citet{furtak2024} & 0.7\\
    4 &CEERS 1670 & $5.24$ & $7.11^{+0.12}_{-0.16}$ & H$\alpha$  & - & -4.97 & \citet{kocevski2023} & -2.7 \\
    5 & JADES1146115 & $6.68$ & $8.61^{+0.27}_{-0.24}$ & H$\alpha$ & - & -5.36 & \citet{juodzbalis2024} & 2.3 \\
    6 & GN-z11 & $10.6$ & $6.2^{+0.3}_{-0.3}$ & N, Ne lines & - & -6.06 & \citet{maiolino2024}, & 1.3\\
     & &  & &  &  & & \citet{fujimoto2023_uncover} & \\
7 & 4286 & $5.84$ & $8.0\pm0.3$ & H$\alpha$ & -4.37 & -4.97 & \citet{greene2024} &-0.6 \\
8 &13123 & $7.04$ & $7.3\pm0.2$ & H$\alpha$ & -4.58 & -4.88 & \citet{greene2024} & 0.5\\
9 &13821 & $6.34$ & $8.1 \pm 0.2$ & H$\alpha$ & -4.37 & -4.97 & \citet{greene2024} & 0.4\\
10 & 20466& $8.5$ & $8.14\pm 0.42$ & H$\alpha$ & -4.88 & -5 & \citet{greene2024}, & 2.9\\ 
&  & & & & & & \citet{kokorev2023} & \\
11 & 23608 & $5.8$ & $7.5 \pm 0.2$ & H$\alpha$ & -5 & -5 & \citet{greene2024} & -1.2\\
12 &35488 & $6.26$ & $7.4 \pm 0.2$ & H$\alpha$ & -4.37 & -4.97 & \citet{greene2024} & -0.5\\
13 &38108 & $4.96$ & $8.4 \pm 0.5$ & H$\alpha$ & -4.37 & -4.97 & \citet{greene2024} & -2.1\\
14 &41255 & $6.76$ & $7.7\pm 0.4$ & H$\alpha$ & -4.58 & -4.88 & \citet{greene2024} & 0.6\\
15 &45924 & $4.46$ & $8.9 \pm 0.1$ & H$\alpha$ & -5 & -5 & \citet{greene2024} & -3.1 \\
16 &CEERS01244 & $4.47$ & $7.5^{+0.04}_{-0.02}$ & H$\alpha$ & -3.04 & -3.64 & \citet{harikane2023bh} & -4.5\\
17 & GLASS 160133& $4.01$ & $6.36^{+0.02}_{-0.01}$ & H$\alpha$ & -3.04 & -3.64 & \citet{harikane2023bh} & -7.3 \\
 18 &GLASS 150029& $4.58$ & $6.56^{+0.03}_{-0.03}$ & H$\alpha$ & -3.04 & -3.64 & \citet{harikane2023bh} & -5.1 \\
 19 &CEERS 00746& $5.62$ & $7.76^{+0.1}_{-0.1}$ & H$\alpha$ & -3.82  & -4.12  & \citet{harikane2023bh} & -1.3\\
 20 &CEERS 01665& $4.48$ & $7.27^{+0.16}_{-0.12}$ & H$\alpha$ & -5.05 & -5.05  & \citet{harikane2023bh} & -4.7\\
 21 &CEERS 00672& $5.66$ & $7.69^{+0.14}_{-0.12}$ & H$\alpha$ & -3.82 & -4.12 & \citet{harikane2023bh} & -1.2 \\
 22 &CEERS 02782& $5.24$ & $7.62^{+0.11}_{-0.11}$ & H$\alpha$ & -3.85 & -3.85 & \citet{harikane2023bh} & -2.2\\
 23 &CEERS 00397 & $6.0$ & $7.0^{+0.25}_{-0.3}$ & H$\alpha$ & -5.45 & -5.45 & \citet{harikane2023bh} & -1.3\\
 24 &CEERS 00717 & $6.93$ & $7.99^{+0.16}_{-0.17}$ & H$\alpha$ & -5.79 & -5.79 & \citet{harikane2023bh} & 1.1 \\
 25 &CEERS 01236 & $4.48$ & $7.25^{+0.2}_{-0.17}$ & H$\alpha$ & -3.04 & -3.64 & \citet{harikane2023bh} & -4.7\\
26 &8083 & $4.64$ & $7.25^{+0.31}_{-0.0.31}$ & H$\alpha$ & -3.06 & -3.90 & \citet{maiolino2023b} & -4.2 \\
 27 &1093& $5.59$ & $7.36^{+0.31}_{-0.31}$ & H$\alpha$ & -3.06 & -3.90 & \citet{maiolino2023b} & -1.7\\
 28 &3608& $5.26$ & $6.82^{+0.38}_{-0.33}$ & H$\alpha$ & -3.83 & -4.52 & \citet{maiolino2023b} & -2.9\\
29 &11836 & $4.40$ & $7.13^{+0.31}_{-0.31}$ & H$\alpha$ & -3.83 & -4.52 & \citet{maiolino2023b} & -5.1\\
30 &20621 & $4.68$ & $7.3^{+0.31}_{-0.31}$ & H$\alpha$ & -3.06 & -3.90 & \citet{maiolino2023b} & -4.0 \\
31 &77652& $5.23$ & $6.86^{+0.34}_{-0.34}$ & H$\alpha$ & -3.06 & -3.90 & \citet{maiolino2023b} & -3.0\\
32 &61888& $5.87$ & $7.22^{+0.31}_{-0.31}$ & H$\alpha$ & -3.83 & -4.52 & \citet{maiolino2023b} & -1.3\\
33 &62309 & $5.17$ & $6.56^{+0.31}_{-0.31}$ & H$\alpha$ & -3.06 & -3.90 & \citet{maiolino2023b} & -3.4 \\
34 &954 & $6.76$ & $7.9^{+0.3}_{-0.3}$ & H$\alpha$ & -3.83 & -4.52 & \citet{maiolino2023b} & 0.8 \\
\\ \hline
    \end{tabular}
    \tablefoot{For each object, we list the ID (column 2), the redshift (column 3), the inferred black hole mass (column 4), the signature (column 5), and the number density from the luminosity function (column 6) as quoted by the cited work (column 8). We assign the same weight to each object in a given luminosity bin for a given work to infer the individual number density calculated in column 7. Finally, we quote the primordial black hole seed mass for each object assuming continuous Eddington accretion onto this seed mass from a redshift of $z=3400$.}
    \label{table1}
\end{table*}

%#################################################################
\section{Inferring the primordial black hole mass function in light of JWST data}
\label{sec_model}
%################################################################# 
Analytically, the time evolution of the black hole mass, $M_{\rm BH}(t) $, can be expressed as 
\begin{equation}
M_{\rm BH}(t) = M_{\rm seed}(t_{\rm seed}) ~ e^{\bigg(\frac{4 \pi G m_p f_{\rm Edd}}{c \sigma_T}\frac{1-\epsilon}{\epsilon} (t-t_{seed}) \bigg)}.
\label{medd}
\end{equation}
Here, $M_{\rm seed}(t_{\rm seed})$ is the seed mass at the starting time of $t_{\rm seed}$, $f_{\rm Edd}$ is the Eddington fraction, $G$ is the universal gravitational constant, $m_p$ is the proton mass, $c$ is the speed of light, $\sigma_T$ is the Thomson scattering cross-section, and $\epsilon$ is the radiative efficiency, for which we use a value of 0.1 throughout the work. We explored the growth of black holes assuming seeding at a redshift of $z_{\rm seed}=25$ (corresponding to $\sim 134$ Myrs after the Big Bang) and seed masses ranging between $10^{2-5}\msun$, and we allowed both Eddington-limited and super-Eddington accretion. With continuous accretion onto the black hole, this formalism implicitly assumes a maximal duty cycle of $100\%$. 

\begin{figure*}
\begin{center}
\center{\includegraphics[scale=0.85]{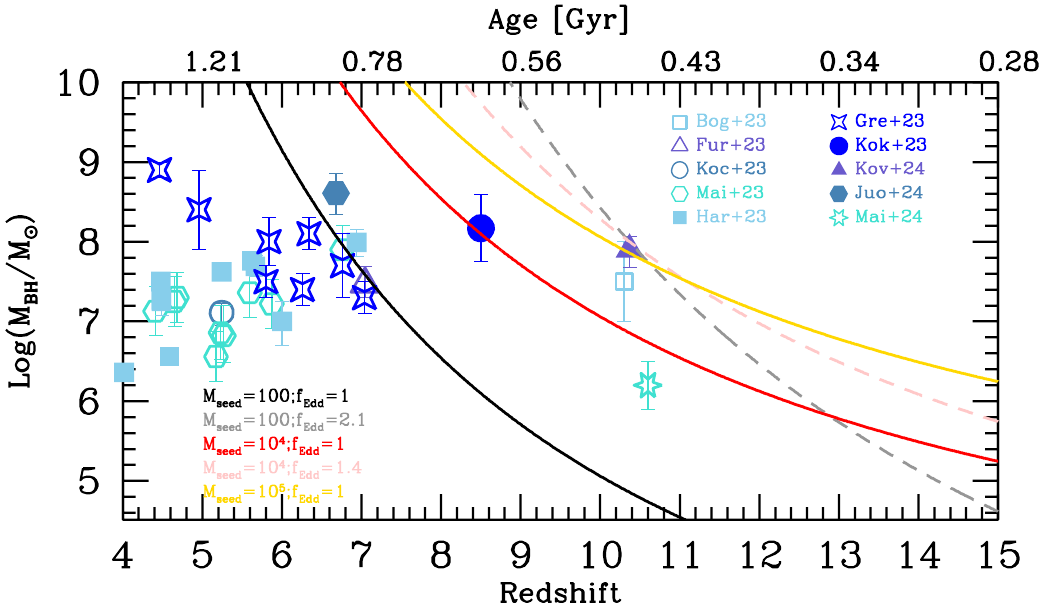}} 
\caption{Growth of the black hole mass as a function of redshift (or cosmic time). Assuming seeding at a redshift of $z_{\rm seed}=25$, we show the black hole growth considering both Eddington-limited ($f_{\rm Edd}=1$) and super-Eddington ($f_{\rm Edd}>1$) accretion for seed masses of $100$, $10^4$, and $10^5\msun$, as marked. As detailed in Table \ref{table1} and marked in the plot, we compare these models to black holes confirmed by \citet{bogdan2024}, \citet{furtak2024}, \citet{kocevski2023}, \citet{maiolino2023b}, \citet{harikane2023bh}, \citet{greene2024}, \citet{kokorev2023}, \citet{kovacs2024}, \citet{juodzbalis2024}, and \citet{maiolino2024}. }
\label{fig_mbhfnz}
\end{center}
\end{figure*}

We then collated data for all black holes confirmed by the JWST at $z \gsim 4$.  As summarised in Table \ref{table1}, a total of 34 such black holes were identified through broad ($\sim 1000-6000~ {\rm Km ~s^{-1}}$) hydrogen balmer lines, which are assumed to trace the kinematics of gas in broad-line regions \citep{kokorev2023, kocevski2023, harikane2023bh, maiolino2023b, furtak2024, greene2024}, through high-ionization lines of nitrogen and neon \citep{maiolino2024}, or through X-ray counterparts in deep {\it Chandra} observations \citep{bogdan2024, kovacs2024}. These were used to infer the existence of massive black holes as large as $10^{7.5}-10^{8.1}\msun$ at $z \gsim 8.5$, only 600 million years after the Big Bang. A key caveat, however, is that these observations apply locally calibrated single-epoch relations to infer black hole masses from observed spectra/X-ray luminosities. This is driven by two key reasons: The first reason is that this provides a redshift-independent estimate of the black hole mass. The second reason is that local relations between line broadening (or X-ray emission) and the black hole mass are associated with small-scale ($<$parsec scale) physics and dynamics that are not expected to show any significant dependence on the redshift \citep[although see][]{king2024, lupi2024}. 

Fig. \ref{fig_mbhfnz} shows that continuous Eddington-limited accretion onto low-mass ($100\msun$) seeds starting at $z_{\rm seed}=25$ is able to yield black hole masses of $M_{\rm BH}\sim 10^{4.8}~ (10^{7.6})\msun$ by $z \sim 10.3 ~ (7)$. As a result, this model is able to explain 26 of the 34 JWST-observed black holes, leaving only 6 outliers. These comprise 2 of the $z \sim 7$ massive black holes \citep{harikane2023bh, juodzbalis2024} and all objects observed at $z \gsim 8.5$ \citep{kokorev2023,bogdan2024,kovacs2024,maiolino2024}. An explanation of these last 6 outliers with low-mass seeds requires invoking continuous super-Eddington accretion with $f_{\rm Edd}=2.1$ that allows assembling $M_{\rm BH}\sim 10^{8}\msun$ as early as $z \sim 10.3$.

An intermediate starting seed mass of $10^4 \msun$ at $z_{\rm seed} =25$ allows us to assemble a mass of $M_{\rm BH}\sim 10^{6.8}~ (10^{9.6})\msun$ by $z \sim 10.3 ~ (7)$, which encompasses {all} of the observed black holes, except for the two highest-redshift objects at $z \gsim 10$ 
\citep{bogdan2024,kovacs2024}. Explaining these with such massive seeds requires invoking $f_{\rm Edd}=1.4$. Finally, already assembling a mass of $M_{\rm BH}\sim 10^{7.9}~ (10^{9.6})\msun$ by $z \sim 10.3 ~ (7)$, continuous Eddington accretion onto massive ($10^5\msun$) seeds is able to explain all of the current observations, precluding the need for super-Eddington accretion. 

\begin{figure*}[h]
\begin{center}
\center{\includegraphics[scale=1.01]{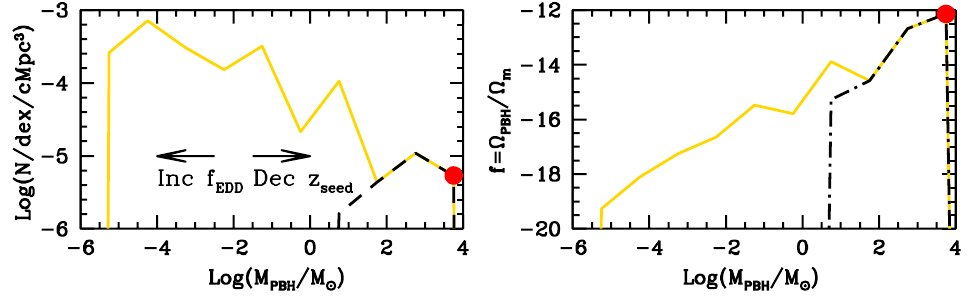}} 
\caption{PBH mass function as a function of the primordial black hole mass ({\it left panel}) and the fraction of dark matter composed of these objects ({\it right panel)}. In both panels, these estimates are obtained assuming Eddington-limited accretion onto all black holes observed by the JWST (solid line; upper limit) and for those that cannot be explained by Eddington-limited accretion onto low-mass seeds (dashed line) and intermediate-mass seeds (point). As shown by arrows in the left panel, super-Eddington accretion would shift the mass function towards lower masses. Decreasing the seeding redshift would shift the mass function to higher masses.}
\label{fig_pbhmf}
\end{center}
\end{figure*}

To summarise, Eddington-limited accretion onto low- and intermediate-mas seeds results in 6 and 2 outliers (out of 34 objects), respectively. Seeds that start at a high mass of $10^5\msun$ at $z=25$ can explain all of the current data. 

We then carried out a {\it Gedankenexperiment} in which we assumed a primordial origin for black holes and calculated the PBH mass assuming that {\it (i)} all of the black holes observed by JWST to have been seeded primordially. This yields the upper limit to the JWST-motivated PBH mass function.  {\it (ii)} We then assumed that only the (6) BHs that cannot be explained by Eddington accretion onto low-mass seeds are primordial, and we assumed {\it (iii)} that only the (2) BHs that cannot be explained by Eddington accretion onto intermediate-mass seeds are primordial. For each of these cases, we calculated the PBH seed mass ($M_{\rm PBH}$) by reversing Eq. \ref{medd} assuming a redshift of $z=3400$. We reasonably ignored any growth of PBHs before this era of matter-radiation equality. Calculating the PBH mass function also requires a number density associated with each seed. We used the number densities quoted in the literature for this (see Table \ref{table1}). These numbers were usually obtained from the luminosity function (Column 6). When several objects contributed to a given luminosity function bin, we assigned each of them the same weight in order to derive the number density per object (Column 7), which are the values used here. The PBH mass function for each of these cases is shown in Fig. \ref{fig_pbhmf} with the PBH seed masses for each object noted in Column 9 of \ref{table1}. 

\begin{figure*}
\begin{center}
\center{\includegraphics[scale=1.1]{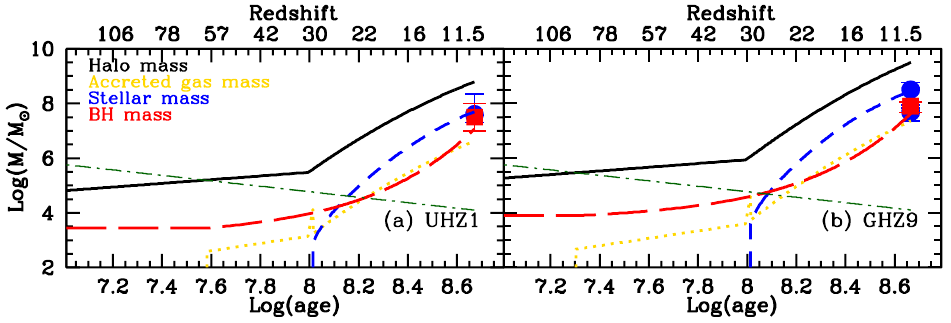}} 
\caption{Assembly of early galaxies in the scenario in which primordial black holes act as the seeds of (isolated) structure formation. The left and right panels show the inferred stellar (blue points) and black hole (red points) masses for UHZ1 and GHZ9, observed at $z \sim 10.3-10.4$. As marked, we show the assembly of the black hole mass (long-dash line), the halo mass (solid line), the accreted gas mass (dotted line), and the stellar mass (short-dash line).  In each panel, the dot-dashed line shows the evolution of the minimum halo mass that is able to bind haloes with a baryonic over-density value of $\delta_b=200$. }
\label{fig_assembly}
\end{center}
\end{figure*}

%We also show the associated PBH formation scale, $\kappa$, as \citep{choudhury-mazum2014}
%\begin{equation}
%\kappa = \frac{\sqrt \gamma}{5.54\times 10^{-24}} \bigg(\frac{M_{\rm PBH}}{1 {\rm gm}} \bigg)^{-1/2} \bigg(\frac{g_*}{3.36}\bigg)^{-1/6},
%\end{equation}
%where \pd{explain terms}.

Case (i) above yields an extended PBH mass function that range between $10^{-5.25}-10^{3.75}\msun$, reflecting the range of black hole masses and redshifts measured by the JWST. The mass function decreases by more than two orders of magnitude from a value of about $10^{-3.1} {\rm cMpc^{-3}}$ at $M_{\rm PBH} \sim 10^{-4.25}\msun$ to $10^{-5.3} {\rm cMpc^{-3}}$ at $M_{\rm PBH} \sim 10^{3.75}\msun$. This reflects the few outliers in terms of mass and redshift as opposed to a large number of reasonable-mass black holes detected at $z \lsim 7$. The massive end of the PBH mass function (for $M_{\rm PBH} \gsim 10^{0.75}\msun$) is completely dominated by the outliers in case (ii) above, that is, those lying above the Eddington-accretion-limited mass onto low-mass seeds. Finally, given their similar masses and redshifts, the outliers in case (iii) above yield PBHs of very similar masses, resulting in a single number density value of $10^{-5.3} {\rm cMpc^{-3}}$ at $M_{\rm PBH} \sim 10^{3.75}\msun$. We note that assuming super-Eddington accretion would shift the mass function towards lower masses. Conversely, decreasing the seeding redshift would shift the mass function to higher masses.

We also calculated the density parameter for PBHs as a function of their mass and converted it into the usual notation of the fraction of dark matter in the form of PBHs as $f = \Omega_{\rm PBH}/\Omega_m$. We show these results in the right panel of Fig. \ref{fig_pbhmf}. We find that even in the maximum scenario in which all JWST black holes are assumed to be primordial, the fractional PBH density parameter reaches $\sim 10^{-19}$ for $M_{\rm PBH} \sim 10^{-5.25}\msun$, which increases to a maximum of $\sim 10^{-12}$ for $M_{\rm PBH} \sim 10^{3.75}\msun$. For this mass range, observational bounds mostly come from microlensing, gravitational waves, accretion, and dynamical effects \citep{kavanagh2024, carr-green2024}, and they typically allow values of $f \lsim 10^{-2}-10^{-3}$. Being ten orders of magnitude lower, our results agree with current constraints, and as expected, they show that JWST-detected black holes contribute very little to the dark matter content. 

%#################################################################
\section{Primordial black holes as seeds of early galaxy assembly}
\label{sec_theory}
%################################################################# 
We now carry out illustrative calculations to show that the PBH masses calculated for the two outliers (UHZ1 and GHZ9) in case (iii) above (requiring super-Eddington accretion onto $10^4\msun$ seeds) could evolve into the systems observed by the JWST. 

%#################################################################
\subsection{Theoretical model}
\label{sec_assembly}
%################################################################# 
In the seed effect, a PBH of mass $M_{\rm PBH}$ at redshift $z_h$ can bind a dark matter halo mass ($M_h$) equal to \citep{carr-silk2018}
\begin{equation}
M_h = \frac{z_{mreq}}{z_h} M_{\rm PBH}.
\label{mhpbh}
\end{equation}
Here, $z_{mreq}=3400$ is the redshift of matter-radiation equality, which is when we assume that the black hole can start to assemble a halo around itself. In this formalism, at redshift $z = 30$, the halo mass is higher by two orders of magnitude than the black hole. Providing the dominant potential, at this point, we assume that the halo transitions grow by smoothly accreting dark matter from the intergalactic medium (IGM). We use the results from high-resolution N-body (dark matter) simulations for this, in which the average halo accretion rate evolves with redshift as \citep{trac2015}
\begin{equation}
\langle \dot M_h \rangle = 0.21 \bigg(\frac{M_h}{10^8 \msun}\bigg)^{1.06} \bigg(\frac{1+z}{7}\bigg)^{2.5} \bigg[\frac{\msun}{\rm yr}\bigg]
\label{mhacc}
\end{equation}  
We caution that this relation was derived for haloes ranging between $10^{8-13}\msun$ at $z \sim 6-10$. However, we assumed that this relation holds out to $z \sim 30$ given the lack of numerically derived results at these early epochs and just used the average relation (without any scatter) for simplicity.

\begin{figure*}
\begin{center}
\center{\includegraphics[scale=1.1]{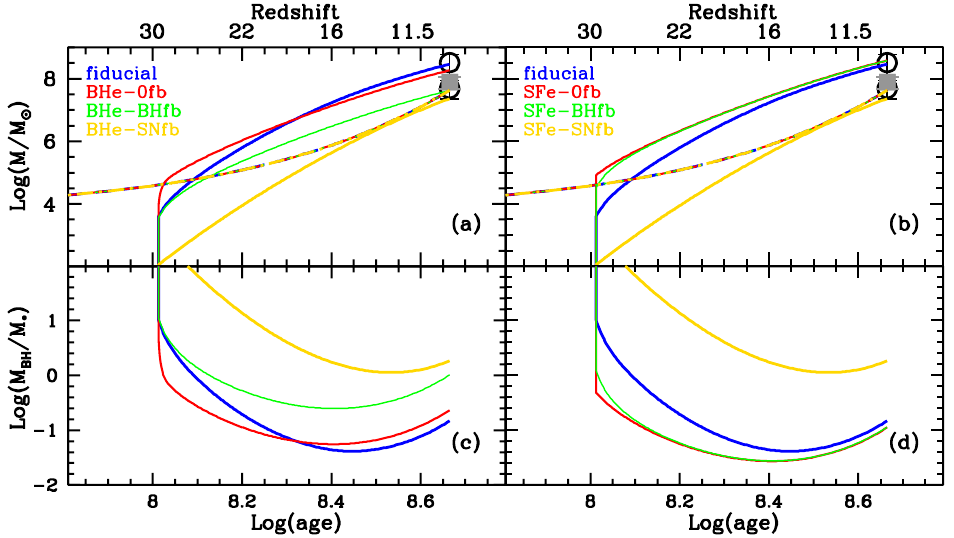}} 
\caption{Assembly of GHZ9 with primordial black holes acting as the seeds of structure formation. In the {top row} (panels a and b), we show the assembly of the stellar mass (solid lines) and black hole mass (dashed lines) for three different models exploring the parameter space (see Table \ref{table2}). The {\it bottom row} (panels c and d) shows the ratios of black hole to stellar mass for each model. While the {\it left} panel shows results for different feedback models in the case where black hole accretion is more efficient (BHe-) than star formation, the {\it right} panel shows results for the cases where star formation is allowed to be more efficient (SFe). The second part of the label in each panel shows the key feedback source. {\it 0fb} stands for no feedback, {\it BHfb} for feedback from black holes alone, and {\it SNfb} for SNII feedback alone. }
\label{fig_paramex}
\end{center}
\end{figure*}

According to the spherical top-hat collapse model for halo formation, the virial temperatures of bound haloes evolve as \citep{barkana-loeb2001}
\begin{equation}
T_{vir} = 1.98\times 10^4 \bigg(\frac{\mu}{0.6}\bigg) \bigg(\frac{M_h}{10^8 h^{-1} \msun}\bigg)^{2/3} \bigg(\frac{\Omega_m \Delta_c}{\Omega_m^z 18 \pi^2}\bigg) \bigg(\frac{1+z}{10}\bigg) {\rm K},
\label{tvir}
\end{equation}
where we used a mean molecular weight value of $\mu = 1.22$ as appropriate for neutral primordial gas, $\Delta_c = 18\pi^2 + 82d-39d^2$, where $d = \Omega_m^z-1$. Finally, $\Omega_m^z = [\Omega_m (1+z)^3][\Omega_m (1+z)^3+\Omega_\Lambda]^{-1}$. At the high redshifts ($z \gsim 6$) considered in this work, $\Omega_m^z \approx 1$, that is, $d =0$ and $\Delta_c = 18\pi^2$. We then modelled the response of baryons to dark matter potential wells \citep[we refer the reader to Sect. 3.2 of the excellent review by][]{barkana-loeb2001}. At $z \lsim 160$, the gas temperature is decoupled from that of the cosmic microwave background (CMB) and evolves as $ \bar T \approx 170[(1 + z)/100]^2 {\rm K}$ \citep{barkana-loeb2001}. Furthermore, the over-density of baryons in a virialised object whose gravitational potential is driven by cold dark matter particles can be written as  
\begin{equation}
\delta_b = \frac{\rho_b}{\bar \rho_b} -1 = \bigg(1+\frac{6 T_{vir}}{5 \bar T}\bigg)^{3/2},
\label{deltab}
\end{equation}
where $\rho_b$ is the gas density, and the bar values denote background quantities. From Eqs. \ref{tvir} and \ref{deltab}, we can calculate 
the minimum halo mass ($M_h^{minb}$) that can host gas with a baryonic over-density of $\delta_b = 200$ at any redshift. We caution that this solution is approximate because it assumes gas to be stationary throughout the object and ignores entropy production in the virialisation shock.

When the assembling halo exceeded the $M_h^{minb}$ value, we allowed it to start building its gas content assuming that the assembling halo is able to drag in gas at a cosmological baryon-to-dark matter fraction, that is, the gas mass assembles at a rate of $\dot M_g = (\Omega_b/\Omega_m) \dot M_h$. We used a redshift step of $\Delta z =0.1$ to obtain the total halo and gas mass that are accreted in any given redshift step (or corresponding time-step). We allowed the central black hole to grow by accreting a fraction of this gas, 
\begin{equation}
\Delta M_{\rm BH} = min[M_{\rm Edd},f_{\rm BH} M_g^i].
\end{equation}
The black hole can therefore grow by accreting the minimum between the Eddington mass and a fraction ($f_{\rm BH}$) of the available gas mass at the start of the redshift step ($M_g^i$). 

We finally allowed star formation to take place at $z=30$. The gas that was left over after black hole accretion was allowed to form stars with an efficiency that is limited by feedback from Type II supernova (SNII; exploding stars more massive than $8\msun$), such that \citep[see][]{dayal2014a, mauerhofer2023}
\begin{equation}
\Delta M_* = (M_g^i-\Delta M_{\rm BH}) \feff,
\end{equation}
where $\feff = min[f_*^{ej}, f_*]$ is the effective star formation efficiency. Physically, this is the minimum between the star formation efficiency that produces enough SNII energy to unbind the remainder of the gas ($f_*^{ej}$) and a maximum star formation efficiency parameter ($f_*$). While $f_*$ is a free parameter, $\fej$ can be calculated for any halo as
\begin{equation}
\fej = \frac{v_c^2}{v_c^2 + f_*^w v_s^2}.
\label{fej}
\end{equation}
Here, $v_c$ is the virial velocity of the host halo, $f_*^w$ is the fraction of SNII energy that couples to the gas component, and $v_s^2 = \nu E_{51} = 611 {\rm km~s^{-1}}$. Here, $\nu = [134 \msun]^{-1}$ is the number of SNII per stellar mass formed for our IMF, and each SNII was assumed to impart an explosion energy of $E_{51}=10^{51}{\rm erg}$. We assumed black hole and SNII feedback to act instantaneously on the gas component that was left after black hole accretion and star formation, leaving a final gas mass at the end of a redshift step of
\begin{equation} 
M_g^f = (M_g^i-\Delta M_{\rm BH} - \Delta M_*) \bigg(1-\frac{E_{ej}}{E_{bin}}\bigg) \bigg(1-\frac{\fej}{\feff}\bigg),
\end{equation}
where the second and third terms on the right-hand side account for black hole and SNII feedback, respectively. Here, $E_{ej} = f_{\rm BH}^w \epsilon \Delta M_{\rm BH} c^2$ is the ejection energy provided by the accreting black hole, $E_{bin}$ is the binding energy of gas in the halo \citep[for details see][]{dayal2019}, and $f_{\rm BH}^w$ is the fraction of black hole energy that couples to gas. This final gas mass in a given redshift step was added to the accreted gas mass to form the new initial gas mass for the next redshift step. Because we assumed black hole and SNII feedback to act instantaneously, the order of star formation or black hole accretion has no impact on the results. 

The free parameters of the model (summarised in Table \ref{table2}) are:the fraction of gas that is accreted onto the black hole ($f_{\rm BH}$), the threshold star formation efficiency ($f_*$), and the fraction of black hole and SNII energies that couple to gas ($ f_{\rm BH}^w $ and $f_*^w$). In the interest of simplicity, in our {fiducial model}, we assumed extremely weak feedback and used $f_{\rm BH}^w=f_*^w=10^{-3}$. 
We find that black holes can always accrete at the Eddington rate, rendering $f_{\rm BH}$ a redundant parameter. We then find the value of $f_*$ by 
matching to stellar mass measured for UHZ1 \citep{bogdan2024} using the PBH mass calculated in Sect. \ref{sec_model} and detailed in Table \ref{table1}: We find a value of $f_*=0.7$, as shown in the left panel of Fig. \ref{fig_assembly}. For this object, we started with a PBH seed mass of $10^{3.4}\msun$ at $z=3400$, which initially binds the same mass in dark matter. By $z=30$, the halo assembling around the black hole reaches a mass of $M_h \sim 10^{5.5}\msun$. Only at $z \sim 60$ is the halo is large enough to exceed  $M_h^{minb}$ and can start assembling its gas content. At this point, the black hole was allowed to start to accrete, and it grew its mass by a factor of 4 between $z =60$ and 30. As a result, by $z=30$, this halo had a black hole mass of $M_{\rm BH} = 10^{4}\msun$, a gas mass of $M_g^i \sim 10^{4.1}\msun$, and $M_* =0$. At $z \lsim 30$, the halo transitions to smoothly accreting dark matter from the IGM at a rate given by Eq. \ref{mhacc}, resulting in a halo mass of $10^{8.8}\msun$ by the source redshift of $10.3$. This was accompanied by smooth accretion of IGM gas, which we (reasonably) assumed to have a cosmological ratio of the gas to dark matter. This means that the gas mass tracked the halo mass, as shown in the same figure. We note that in this simplistic model of an isolated halo, accretion is the only mode through which the system can gain gas. In terms of the black hole mass, we find that it is able to grow efficiently at the Eddington rate. By $z =10.3$, this black hole achieves a mass $\sim 10^{7.1}\msun$, comparable to the observationally inferred mass for UHZ1. Finally, given the low halo mass, star formation always proceeds at the ejection efficiency ($\fej$) such that at any time, the only gas mass available is the mass accreted in that time-step. Given the higher efficiency of the process, the stellar mass starts to exceed the black hole mass as early as $z \sim 25.4$ and reaches a final value of $M_* \sim 10^{7.7}\msun$ by the observed redshift. In this case, the final ratio of black hole to stellar mass is about $0.25$.

We then applied the same parameters to GHZ9 (right panel of Fig. \ref{fig_assembly}) as a sanity check. The stellar mass of this object ranges between $M_* = (4.9^{+2.0}_{-2.6}) \times 10^7 \msun$ \citep{atek2023b} and $M_* =  (3.3^{+2.9}_{-2.4}) \times 10^8 \msun$ \citep{castellano2023}. 
This object starts with a $2.8$ times higher primordial black hole with a mass of  $10^{3.9}\msun$ , which allows its halo to exceed $M_h^{minb}$ as early as $z \sim 90$. By the observed redshift of $z =10.4$, this object has a halo mass of $M_h \sim 10^{9.5}\msun$. Additionally, without any change in parameters, we obtain final stellar and black hole mass values of $10^{8.5}$ and $10^{7.6}\msun$, that is, $\mbh/\ms \sim 0.12$, which agrees very well with the observed values of the black hole mass \citep{kovacs2024} and the stellar mass \citep[as inferred by][]{castellano2023}, as shown.

\begin{table*}
\caption{Parameter values.}
\centering
    \begin{tabular}{|c|c|c|c|c|c|c|}
    \hline
   No. & Model name & properties & $f_{\rm BH}$  & $f_*$ & $f_{\rm BH}^w$ & $f_*^w$ \\
    \hline
    1 & fiducial & weak feedback & - & 0.7 & $10^{-3}$ & $10^{-3}$ \\
    2 & BHe-0fb &  BH accretion efficiency>star formation efficiency, no feedback & 0.9 & 0.1 & $0$ & $0$ \\
    3 & BHe-BHfb & BH accretion efficiency>star formation efficiency, BH feedback only &  0.9 & 0.1 & $0.1$ & $0$ \\
    4 & BHe-SNfb &  BH accretion efficiency>star formation efficiency, SN feedback only & 0.9 & 0.1 & $0$ & $0.1$ \\
    5 &SFe-0fb &  star formation efficiency>BH accretion efficiency, no feedback &  0.1 & 0.9 & $0$ & $0$ \\
    6 & SFe-BHfb & star formation efficiency>BH accretion efficiency, BH feedback only &  0.1 & 0.9 & $0.1$ & $0$ \\
    7 & SFe-SNfb &  star formation efficiency>BH accretion efficiency, SN feedback only &  0.1 & 0.9 & $0$ & $0.1$
\\ \hline
    \end{tabular}
    \tablefoot{We explore a wide range of parameter spaces for the models listed  in column 2 and their properties detailed in column 3. For each, we show  the fraction of gas available for accretion onto the black hole (column 4), the maximum threshold efficiency for star formation (column 5), and the fractions of black hole and SNII feedback energies that can couple to gas (columns 6 and 7, respectively). }
    \label{table2}
\end{table*}

% ******************************************************************
\subsection{Impact of free parameters on the model results}
\label{sec_parex}
% *******************************************************************
As detailed in Table \ref{table2}, we explored a wide range of models, including extreme combinations of the gas-mass fractions available for black hole accretion/star formation and  the fraction of black hole/SNII feedback coupling to the gas content. We applied these models to GHZ9 to highlight the impact of the parameter values on the assembled stellar and black hole masses. The results of these calculations are also shown in Fig. \ref{fig_paramex}. We start by noting that the exact order of black hole accretion or star formation has no bearing on our results, and black holes always accrete at the Eddington rate. As a result, the black hole assembly is the same in every model, even though we used different values that spanned the extrema of the parameter space.  

We start by discussing the set of models in which black holes are allowed to accrete the minimum between the Eddington mass and $90\%$ of the available gas mass, with $10\%$ of the gas forming stars at most, at any step. As shown in {\it panel a} of Fig. \ref{fig_paramex}), in the model without feedback (BHe-0fb), the stellar mass reaches a final value of $M_* \sim10^{8.3} \msun$, in agreement with the inferred value from \citet{castellano2023}, that is, $\mbh/\ms \sim 0.23$, as shown in panel (c). When the impact of black hole feedback is included (BHe-BHfb), a stellar mass is reached that is almost five times lower and has a value of $M_* \sim 10^{7.6}\msun$ \citep[in agreement with the mass estimate from][]{atek2023b}. Interestingly, in this case, the model yields a black hole mass that exceeds the stellar mass, with a value of $\mbh/\ms \sim 1.01$ (panel c). Finally, when we consider SNII feedback (BHe-SNfb), a very slow buildup of the stellar mass is caused that reaches a final value of only $M_* \sim 10^{7.4}\msun$, again yielding a black hole mass that exceeds the stellar mass by a factor 1.86, as shown in panel (c).

The case of efficient star formation, in which the minimum between the ejection efficiency up to $90\%$ of the gas is allowed to form stars, is shown in {\it panel b} of the same figure. In this case, in the cases without feedback (SNe-0fb) and with black hole feedback (SNe-BHfb), the stellar mass is only 0.1 dex higher than that from the {fiducial} relation, resulting in $\mbh/\ms \sim 0.1$, as shown in panel d. These stellar masses agree with the higher mass estimates of \citet{castellano2023}. In the SNII feedback case (SNe-SNfb), a combination of high star formation rates and high feedback results in a final stellar mass of $M_* \sim 10^{7.4}\msun$, which agrees with the values inferred by \citet{atek2023b}. This results in $\mbh/\ms \sim 1.86$ (panel d).

To summarise, essentially the entire parameter space we explored is compatible with the black hole and stellar masses inferred for GHZ9 given the wide range of $M_*$ values that are observationally allowed for this object. Interestingly, the ratios of the black hole to stellar mass inferred from the different models we explored range between $0.1-1.86$, which means that in some cases (high supernova feedback), the black hole even grows to be more massive than its host halo. 

%#################################################################
\section{Conclusions and discussion}
\label{sec_conc}
%################################################################# 
The JWST has been instrumental in shedding light on black holes in the first billion years of the Universe. These unprecedented datasets have raised a number of tantalising issues, including an over-abundance of black holes \citep{harikane2023bh, maiolino2023b, greene2024}, the presence of overly huge black holes ($M_{\rm BH} = 10^{6.2-8.1}\msun$) as early as $z \sim 8.5-10.6$  \citep{kokorev2023,maiolino2024,bogdan2024, kovacs2024}, and systems with unprecedentedly high ratios of the stellar to black hole mass of $\mbh/\ms\gsim 30\%$ \citep{kokorev2023,juodzbalis2024,bogdan2024, kovacs2024}, compared to local relations, where these values are lower by two orders of magnitude \citep[e.g.][]{volonteri2016}. 

A number of works have cautioned about the use of standard virial indicators in estimating black hole masses for these systems \citep{king2024}, and others used BLR models to estimate black hole masses that were lower by an order of magnitude \citep{lupi2024}. Finally, it has been noted that selection biases and measurement uncertainties might account for the offset of the inferred high-redshift $\mbh-\ms$ relation \citep{li2024}. These caveats notwithstanding, an enormous body of theoretical work has focused on {astrophysical} solutions, including high-mass seeding mechanisms \citep{bogdan2024, kovacs2024, maiolino2023b, natarajan2024}, episodic super-Eddington accretion phases \citep{schneider2023, maiolino2024}, extremely efficient accretion onto low- or intermediate-mass seeds \citep{furtak2024, dayal2024}, or high-mass seeds that grow through mergers with low-mass seeds and/or efficient accretion \citep{bhowmick2024}. 

We explored a complementary {cosmological} solution provided by PBHs. These are seeded in the early Universe through mechanisms including collapse from scale-invariant fluctuations, cosmic strings, cosmological phase transitions, or a temporary softening of the equation of state \citep[see excellent reviews by e.g.][]{carr2005, carr-green2024}. A key advantage of this scenario is that individual PBHs can act as seeds of structure formation by binding their dark matter halo around themselves \citep{hoyle1966, ryan1972, carr-rees1984, mack2007,carr-silk2018}. 

Starting by collating data on all of the 34 JWST-detected black holes confirmed at $z \gsim 4$ (summarised in Table \ref{table1}), we carried out analytic calculations to estimate the seed masses and accretion rates that would allow their assembly. We used these observations to derive limits on the PBH mass function. We then developed an illustrative model to show a PBH that assembles its dark matter and baryonic content in the seed scenario. This was successfully applied to the highest-redshift outlying black holes observed (UHZ1 and GHZ9). Our key findings are listed below.
\begin{enumerate}
\item Starting at a redshift of $z=25$, continuous Eddington-limited accretion onto low-mass ($100\msun$) seeds is able to explain 26 of the 34 JWST-observed black holes, leaving only 6 outliers (2 at $z \sim 7$ and all of the 4 at $z \gsim 8.5$). Explaining these requires invoking continuous super-Eddington accretion with $f_{\rm Edd}=2.1$. Eddington accretion onto intermediate-mass seeds ($10^4 \msun$) explains 32 of the observed black holes except for the two highest-redshift objects at $z \gsim 10$ \citep{bogdan2024,kovacs2024}. These objects can be explain by invoking $f_{\rm Edd}=1.4$. Continuous Eddington accretion onto massive ($10^5\msun$) seeds is able to explain all of the current observations, precluding the need for super-Eddington accretion. 

\item Assuming a PBH origin for {all} JWST black holes yields the upper limit to the PBH mass function. This shows an extended shape, with number densities $\sim 10^{-3.1} {\rm cMpc^{-3}}$ at $M_{\rm PBH} \sim 10^{-4.25}\msun$ to $10^{-5.3} {\rm cMpc^{-3}}$ at $M_{\rm PBH} \sim 10^{3.75}\msun$. However, even in this maximum scenario, the fractional PBH density parameter has a maximum value $\sim 10^{-12}$, which is lower by ten orders of magnitude than current constraints. This implies that PBHs from these JWST-detected early sources contribute very little to the dark matter content. 

\item Considering PBHs to form the seeds of structure formation, we successfully showed that the two highest-redshift sources (UHZ1 at $z \sim 10.3$ and GHZ9 at $z \sim 10.4$) assemble their dark matter and baryonic contents with the same parameters that govern star formation and black hole accretion and their associated feedbacks. 

\item Exploring a wide swathe of parameter space in terms of black hole gas accretion, star formation efficiency, and their associated feedbacks applied to GHZ9, we find ratios of the black hole to stellar mass between $0.1-1.86$, which means that in some cases (high supernova feedback), the black hole even grows to be more massive than its host halo. 

\end{enumerate}

We end with a few caveats: From the observational side, as noted, black hole masses are inferred using single-epoch measurements calibrated against local relations. Furthermore, the inferred stellar masses intricately depend on the assumed star formation history and stellar population properties (nebular emission inclusion, metallicity, and dust attenuation) and, most crucially, on the initial mass function. Systematic uncertainties arising from these quantities can change the stellar mass estimates by about an order of magnitude \citep{wang2024}. Finally, we note that the black holes that are currently being observed only form a subset of the underlying population, which might induce systematics in the observed relations \citep[e.g.][]{li2024}.

In terms of the theoretical model, we have presented what is mostly an illustrative model of how a black hole can seed a halo and its baryonic components. We note that a number of assumptions were made that must be improved further: We transitioned from a linear (black hole dominated) growth of the halo to non-linear when the halo exceeds the black hole by two orders of magnitude, which must be investigated further. The accretion rates used here were also derived for higher-mass haloes at lower redshifts. Star formation and black hole accretion and their associated feedback could all be modelled in more detail. We also assumed that the black hole is able to remain stable at the centre of the halo and continue to accrete with a duty cycle of 100\%. 

In addition to the JWST, facilities such the Laser Space Interferometer Antenna (LISA) and Athena will be utterly crucial in the next decades for our understanding of black holes and their hosts in the first billion years.

% *********************************************

%***************************************************************************
\begin{acknowledgements}
PD acknowledge support from the NWO grant 016.VIDI.189.162 (``ODIN") and warmly thanks the European Commission's and University of Groningen's CO-FUND Rosalind Franklin program. PD thanks A. Mazumdar for illuminating discussions and L. Furtak, J. Greene and M. Mosbech for their comments that have added crucially to the quality of the work.
\end{acknowledgements}

%****************************************************************************

%%%%%%%%%%%%%%%%%%%%%%%%%%%%%%
%\vspace{-0.8cm}
\bibliographystyle{aa} % style aa.bst
\bibliography{bh}

\end{document}